\documentclass[prc,twocolumn,superscriptaddress,showpacs,preprintnumbers,amsmath,amssymb]{revtex4-1}
\usepackage[dvips]{graphicx}
\usepackage{dcolumn}
\usepackage{bm}

\begin{document}
\title{
Fragmentation of electric dipole strength in $N=82$ isotones
}
\author{
Mitsuru Tohyama}
\affiliation{
Kyorin University School of Medicine, Mitaka, Tokyo 181-8611, Japan}
\author{
Takashi Nakatsukasa}
\affiliation{
RIKEN Nishina Center, Wako, Saitama 351-0198, Japan
}
\affiliation{
Center for Computational Sciences, University of Tsukuba, Tsukuba 305-8571, Japan
}

\begin{abstract}
Fragmentation of the dipole strength in the $N=82$ isotones $^{140}$Ce, $^{142}$Nd and $^{144}$Sm  
is calculated using the second random-phase approximation (SRPA). 
In comparison with the result of the random-phase approximation (RPA),
the SRPA provides the additional damping of the giant dipole resonance and
the redistribution of the low-energy dipole strength.
Properties of the low-energy dipole states are significantly changed
by the coupling to two-particle-two-hole ($2p2h$) states,
which are also sensitive to the correlation among the $2p2h$ states.
Comparison with available experimental data shows
a reasonable agreement for the low-energy $E1$ strength distribution.
\end{abstract}

\maketitle

The low-energy dipole states,
often referred to as the pygmy dipole resonance (PDR),
have attracted recent experimental \cite{zilges,enders,volz,savran,schweng,wieland}
and theoretical interests \cite{yoshida-1,pena,paar-2,ebata,inakura,martini}
(see also the recent review \cite{paar-1} and references therein).
It is also of significant astrophysical interest,
since the low-energy dipole strengths close to the neutron threshold
strongly affect
the astrophysical r-process nucleosynthesis \cite{goriely}.

The quasiparticle random-phase approximation based on the
Hartree-Fock-Bogoliubov ground state (HFB+QRPA)
has been extensively used to study the PDR as well as the giant dipole resonances (GDR).
Recent systematic calculations \cite{yoshida-2} for the Nd and Sm isotopes show 
that although the HFB+QRPA nicely reproduces characteristic features of
the shape phase transition in the GDR,
it fails to produce the low-energy dipole strengths at $E_x=5.5\sim 8$ MeV,
observed in the $N=82$ isotones, $^{142}$Nd and $^{144}$Sm \cite{zilges,volz}.
The disagreement suggests that the coupling to complex configurations,
such as multi-particle-multi-hole states, 
are required to study the PDR in these nuclei.
In fact, the quasiparticle-phonon model (QPM),
which takes into account coupling to multi-phonon states,
successfully reproduces
the low-energy dipole strengths in the $N=82$ nuclei \cite{enders,savran}.
A similar approach based on the relativistic mean-field model
has also been used to study the PDR in the tin and nickel isotopes \cite{Litv}.
These models assume
the multi-phonon characters of the complex states
and violate the Pauli principle.
Thus, it is desirable to study properties of the PDR
with a method complementary to these phonon-coupling approaches.
In this work,
we present studies for
the dipole excitations in the $N=82$ isotones,
with the second random-phase approximation (SRPA) (Ref. \cite{srpa} and
references therein).
The SRPA explicitly incorporates the two-particle-two-hole ($2p2h$) states
instead of ``two-phonon'' states,
and respects the Pauli principle in the $2p2h$ configurations.
Recently, the low-energy dipole states in $^{40,48}$Ca have been studied
with the SRPA \cite{gambacurta11}, which suggests that
the coupling between one-particle-one-hole ($1p1h$) and $2p2h$ configurations
enhances the electric dipole ($E1$) strength in the energy range
from 5 to 10 MeV.
We investigate whether a similar effect can be observed
in the isotones of $N=82$.
Since there are many dipole states with small $E1$ strengths
in the energy region below 8 MeV,
it is difficult to compare property of each state with the experiment.
Thus, we perform the comparison of integrated properties at low energies.

The SRPA equation is written in the matrix form\cite{srpa}
\begin{eqnarray}
\left(
\begin{array}{cc}
a&c\\
b&d
\end{array}
\right)\left(
\begin{array}{c}
x^\mu\\
X^\mu
\end{array}
\right)
=\omega_\mu
\left(
\begin{array}{c}
x^\mu\\
X^\mu
\end{array}
\right),
\label{ERPA}
\end{eqnarray}
where $x_{ph}^\mu$ and $X_{pp'hh'}^\mu$ ($p\leftrightarrow h$)
are the $1p1h$ and $2p2h$ transition
amplitudes for an excited state with an excitation energy $\omega_\mu$.
The explicit expression for the matrices $a$, $b$, $c$, and $d$
are given in Ref. \cite{Toh}.

The Skyrme interaction of the SIII parameter set
is used to calculate the Hartree-Fock
single-particle states.
The continuum states are discretized by confining the single-particle wave functions in a sphere of radius of 20 fm. 
Single-particle states with the angular momenta $j_\alpha \le 15/2$
up to 30 MeV in energy ($\epsilon_\alpha < 30$ MeV)
are adopted for the $1p1h$ space ($x_{ph}^\mu$ and $x_{hp}^\mu$),
both for protons and neutrons.
This roughly amounts to one hundred single-particle states.
For the $2p2h$ amplitudes ($X_{pp'hh'}^\mu$ and $X_{hh'pp'}^\mu$),
we truncate the space into the one made of the single-particle states
near the Fermi level,
the $2p_{3/2}$, $2p_{1/2}$, $1g_{9/2}$, $1g_{7/2}$, $2d_{5/2}$, $2d_{3/2}$, $3s_{1/2}$, $1h_{11/2}$, and $1h_{9/2}$ orbits 
for protons and the
$2d_{5/2}$, $2d_{3/2}$, $1h_{11/2}$, $1h_{9/2}$, $2f_{7/2}$, and $1i_{13/2}$ orbits for neutrons.
The proton orbits up to the $1g_{7/2}$ orbit are assumed to be
fully occupied in the ground state of $^{140}$Ce,
while the proton $2d_{5/2}$ orbit
is to be partially occupied in the ground states of $^{142}$Nd and $^{144}$Sm.
The numbers of $1p1h$ and $2p2h$ amplitudes in the SRPA are
about 800 and 9000, respectively.

For calculation of the SRPA matrix elements,
we employ a residual interaction of the $t_0$ and $t_3$ terms of the SIII
interaction.
Since the residual interaction is not fully consistent with the one
used in the calculation of the single-particle states,
it is necessary to adjust the strength of the residual interaction
so that the spurious mode corresponding
to the center-of-mass (COM) motion comes at zero excitation energy in the RPA.
This condition determines
the renormalization factor $f$ for the residual interaction
($t_0\rightarrow f \times t_0$ and $t_3\rightarrow f \times t_3$).
We obtain $f=0.73$ for $^{142}$Nd, and similar values for
other nuclei as well.
Since the coupling between the spurious COM motion and $2p2h$ configurations
is weak, 
these renormalization factors may
approximately produce zero energy in the SRPA as well.
Thus, we use this interaction
for the calculation of the matrices $a$, $b$, and $c$
in Eq. (\ref{ERPA}).
For the residual interaction for the matrix $d$,
following a prescription in Ref.~\cite{Toh},
we introduce a zero-range interaction
$v_0\delta^3({\bm r}-{\bm r'})$ in addition to
the original $t_0$ and $t_3$ terms,
then, fix the parameter $v_0$ by approximately reproducing
the excitation energy of the lowest $1^-$ state in $^{142}$Nd
($v_0=-570$ MeV fm$^3$). 
With these residual interactions in the given model space,
the spurious mode appears at a small imaginary energy ($\omega^2\approx -1$
 MeV$^2$) in the SRPA.

\begin{figure}[t]
%\begin{center} 
\includegraphics[width=0.50\textwidth]{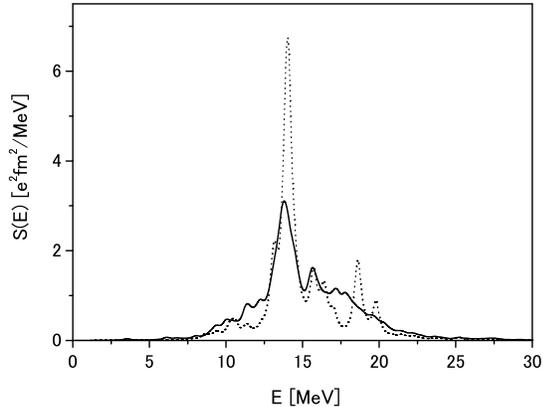}
%\end{center}
\caption{Strength functions calculated in the SRPA (solid line) and RPA (dotted line) for $^{140}$Ce.
An artificial width $\Gamma=0.5$ MeV is used for smoothing.
See text for details.}
\label{fig1} 
\end{figure}
\begin{figure}[tb]
%\begin{center} 
\includegraphics[width=0.50\textwidth]{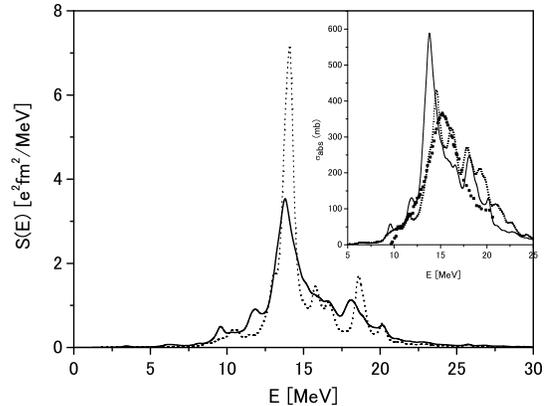}
%\end{center}
\caption{Same as Fig. \ref{fig1} but for $^{142}$Nd.
In the inset the total photoabsorption cross section
calculated from the strength function in the SRPA (solid line) is compared with the experimental data (dots) \cite{carlos}.
The dotted line in the inset denotes the result calculated with $f=0.9$.} 
\label{fig2} 
\end{figure} 
\begin{figure}[tb]
%\begin{center} 
\includegraphics[width=0.50\textwidth]{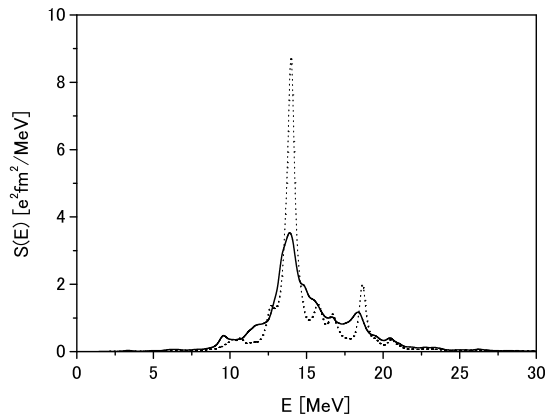}
%\end{center}
\caption{Same as Fig. \ref{fig1} but for $^{144}$Sm.} 
\label{fig3} 
\end{figure}

We first show the results for the GDR.
The $E1$ strength functions,
$S(E)\equiv \sum_n |\langle n| r Y_{1\mu} |0\rangle|^2 \delta(E-E_n)
= dB(E1;1^-\rightarrow 0^+_{\rm gs})/dE$,
calculated in the SRPA (solid line) and RPA (dotted line) for $^{140}$Ce, $^{142}$Nd, and $^{144}$Sm 
are shown in Figs. \ref{fig1}, \ref{fig2} and \ref{fig3}, respectively.
We use the $E1$ operator with the recoil charges,
$Ne/A$ for protons and $-Ze/A$ for neutrons,
for the calculation of $S(E)$.
The obtained discrete strength functions are smoothed with a small width
($\Gamma=0.5$ MeV) of Lorentzian.
The energy-weighted strength summed up to 50 MeV exhausts 87\% of 
the energy-weighted sum-rule value including the enhancement term
arising from the momentum-dependent parts of the Skyrme
interaction.
The strength distributions of the GDR in the SRPA are broadened,
compared to the RPA, due to the coupling to the $2p2h$ states.
In the inset of Fig. \ref{fig2},
the total photoabsorption cross section (solid line)
calculated in the SRPA is compared with the experimental data \cite{carlos}.
The shape of the GDR depends on the parameter $f$, 
whereas it is little affected by the parameter $v_0$.
The GDR peak position and the profile are better described by a slightly
larger value of $f$ (See the dotted line in the inset of
Fig. \ref{fig2}).
Our calculation indicates that the coupling to the $2p2h$ induces
an additional broadening due to the spreading width, however,
the peak position is close to that obtained in the RPA calculation.
This is very different from the recent SRPA calculation for
$^{16}$O in Ref.~\cite{gambacurta10},
which indicates a large shift of the GDR peak energy (more than 5 MeV)
but almost no broadening.
At present, we do not fully understand the origin of this discrepancy.
More quantitative analysis of the GDR require
an improvement of the present calculation, especially,
a self-consistent treatment of the residual interaction and
the enlargement of the $2p2h$ space.

\begin{figure} 
\begin{center} 
\includegraphics[height=6cm]{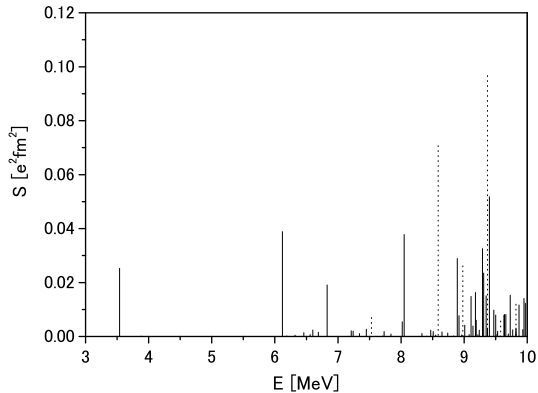}
\end{center}
\caption{Low-energy $E1$ strength distributions,
$B(E1;1^- \rightarrow 0^+_{\rm gs})$
calculated in the SRPA (solid line) and RPA (dotted line) for $^{140}$Ce.
} 
\label{fig4} 
\end{figure}
\begin{figure} 
\begin{center} 
\includegraphics[height=6cm]{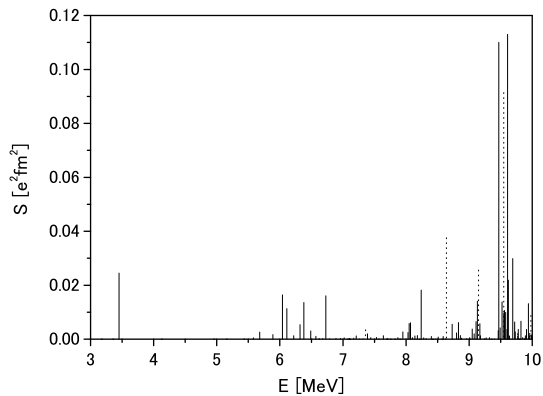}
\end{center}
\caption{Same as Fig. \ref{fig4} but for $^{142}$Nd.} 
\label{fig5} 
\end{figure} 
\begin{figure} 
\begin{center} 
\includegraphics[height=6cm]{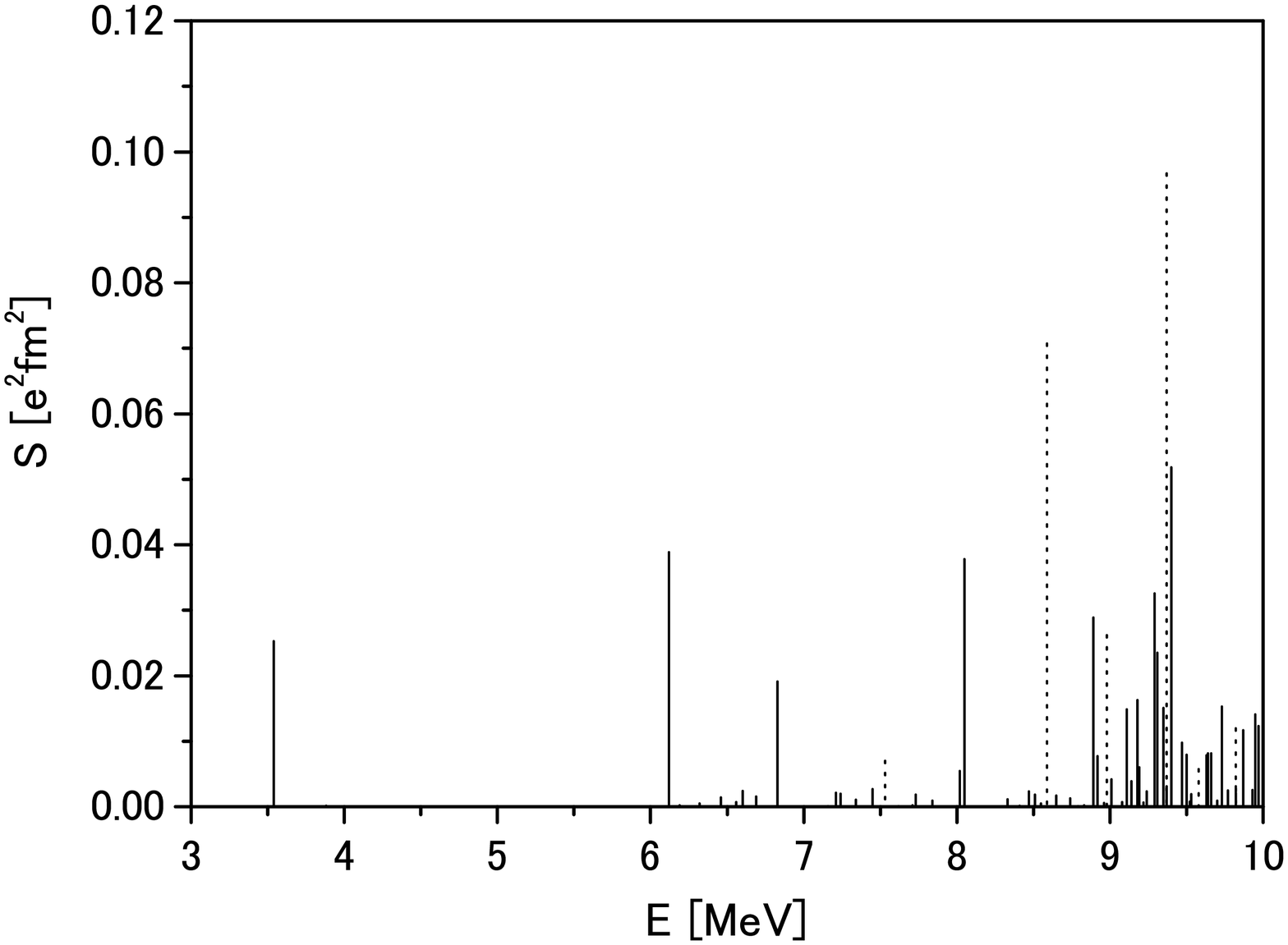}
\end{center}
\caption{Same as Fig. \ref{fig4} but for $^{144}$Sm.} 
\label{fig6} 
\end{figure}
\begin{table}[t]
\caption{Mean energies $\bar{E}$ and
 summed $B(E1)\uparrow$ values for the low-energy
dipole states. The experimental values are taken from Ref. \cite{volz}.
See text for details.}
\begin{center}
\begin{tabular}{c|ddd|ddd} \hline
 &\multicolumn{3}{c|}{$\bar{E}_x$ [MeV]}&\multicolumn{3}{c}{$\sum B(E1)\uparrow$ [$e^2$fm$^2$]}\\ \hline 
Nucleus & \multicolumn{1}{c}{RPA} & \multicolumn{1}{c}{SRPA} & \multicolumn{1}{c|}{Exp} & \multicolumn{1}{c}{RPA} & \multicolumn{1}{c}{SRPA} & \multicolumn{1}{c}{Exp} \\ \hline
$^{140}$Ce & 7.53 & 6.47 & 6.28 & 0.021 & 0.219 & 0.308 \\
$^{142}$Nd & $-$ & 6.31 & 6.07 & 0.0 & 0.224 & 0.184 \\
$^{144}$Sm & $-$ & 6.04 & 5.69 & 0.0 & 0.233 & 0.208 \\ \hline
\end{tabular}
\label{tab2}
\end{center}
\end{table}
Next, let us discuss the low-energy $E1$ strengths.
In contrast to the GDR at high energy,
the truncation of the $2p2h$ configurations is supposed to be less serious.
The $E1$ strengths, $B(E1)\downarrow$,
below 10 MeV in $^{140}$Ce, $^{142}$Nd, and $^{144}$Sm are shown
in Figs. \ref{fig4},
\ref{fig5}, and \ref{fig6}, respectively.
In the RPA calculation,
there is very little $E1$ strength in the energy region below
8 MeV, which agrees with the result of the QRPA calculation \cite{yoshida-2}.
However, this is different from the experimental findings \cite{zilges,enders,volz,savran}.
In the SRPA calculation,
the coupling to the $2p2h$ configurations leads to
a considerable $E1$ strength in this energy region.
To make a quantitative comparison with experiment,
the mean excitation energies and the summed $B(E1)\uparrow$ values
for low-energy dipole states are calculated in the same way as the
experiment \cite{volz}:
The mean energy is defined as $\bar{E}\equiv \sum EB(E1)/\sum B(E1)$,
in which the summation is performed for
the dipole states below 7.7 MeV for $^{140}$Ce,
those below 7.1 MeV for $^{142}$Nd, and
below 7.0 MeV for $^{144}$Sm.
The lowest $1^{-}_1$ states are excluded in the summation.
The result is tabulated in Table \ref{tab2}.
For comparison, the RPA values, which include the lowest $1^-$ state,
are listed in the table, but
no $1^-$ state is predicted below 7.1 MeV for $^{142}$Nd and $^{144}$Sm.
Although the calculated mean energies are slightly larger than the observed values, their isotone dependence is
consistent with the experiment and the summed transition probabilities are comparable to the experimental values \cite{volz}.

In the RPA calculation,
the neutron excitations are dominant in the low-lying states \cite{paar-1}.
The present RPA calculation also indicates,
for instance in $^{142}$Nd, that
the largest components of the low-lying dipole states located
at $E_x=7.36$, 8.64, 9.15,
and 9.55 MeV are $(2p_{1/2}\rightarrow 2d_{3/2})\pi$,
$(3s_{1/2}\rightarrow 3p_{3/2})\nu$, $(3s_{1/2}\rightarrow 3p_{3/2})\nu$, and $(3s_{1/2}\rightarrow 3p_{1/2})\nu$,
respectively. 
In the SRPA,
we see a significant fragmentation of the dipole strength into the
energy range of $5<E<8$ MeV, in addition to the emergence of
the lowest $1^-_1$ state at $E\approx 3.5$ MeV.
Many of these low-lying dipole states consist of proton $2p2h$ characters,
such as
$([1g_{7/2} 2d_{5/2}]^{6^+} \rightarrow [1h_{11/2} 2d_{3/2}]^{7^-})\pi$ and
$([1g_{7/2} 2d_{5/2}]^{6^+} \rightarrow [1h_{11/2} 3s_{1/2}]^{5^-})\pi$.
These proton $2p2h$ configurations come down to the lower energy
because of the coupling to
the $2p2h$ configurations consisting of the neutron $1p1h$ transition
from the $1h_{11/2}$ orbit to the $1h_{9/2}$ orbit and
the proton $1p1h$ transitions
from the $1g_{7/2}$ orbit (or $2d_{5/2}$ orbit) to the $1h_{11/2}$ orbit;
$\pi 1g_{7/2} \nu 1h_{11/2} \rightarrow \pi 1h_{11/2} \nu 1h_{9/2}$ and.
$\pi 2d_{5/2} \nu 1h_{11/2} \rightarrow \pi 1h_{11/2} \nu 1h_{9/2}$ and.
We have confirmed the importance of these proton-neutron $2p2h$ configurations
by performing the SRPA calculation in a smaller $2p2h$ space.
The SRPA calculation with the neutron $1h$ orbits qualitatively produces
the same result.

Finally, let us discuss the property of the lowest $1^-$ state.
The excitation energies and the reduced transition probabilities $B(E1)\uparrow$ of 
the $1^{-}_1$ states in $^{140}$Ce, $^{142}$Nd and $^{144}$Sm are compared with the experimental values \cite{volz}
in Table \ref{tab1}. The calculated 
excitation energies decrease with increasing proton number, which is consistent with the experiment. However,
the SRPA calculations overestimate the $B(E1)\uparrow$ values by a factor of $2.7 - 3.5$. 
The structure of the $1^-_1$ states in these nuclei is supposed to be
predominantly of the two-phonon 
quadrupole-octupole character $2^+\otimes3^-$ \cite{zilges,volz}.
However, in the present SRPA calculation,
the  $2p2h$ configuration $([\pi 1g_{7/2} \nu 1h_{11/2}]^{1^-} \rightarrow
[\pi 1h_{11/2} \nu 1h_{9/2}]^{2^+})$ is dominant in these $1^-_1$ states,
which differs from the two-phonon $2^+\otimes3^-$ character.
The pairing correlation, which is not taken into account in the present
calculation, may play an important role
for a better description of the two-phonon character of the $1^-_1$ states,
because they are essential in the description of the lowest quadrupole and
octupole states.
Furthermore, it has been known that the SRPA fails to describe the collectivity
of the two-phonon states \cite{tamura}.
This is due to the fact that the next-leading terms in the two-phonon
state are missing in the SRPA.
These missing terms beyond the SRPA can be taken into account
by introducing $X_{php'h'}$ amplitudes in Eq. (\ref{ERPA}).
A general equation for the extended RPA formalism with the ground-state
correlation is given in Ref. \cite{ts07}.
Another possible method to improve the description of the two-phonon states
is the dressed-four-quasi-particle approach proposed in Ref. \cite{kanesaki}.
These are beyond the scope of the present work, but of
significant interest in future.

\begin{table}[t]
\caption{Excitation energies $E$ and
$B(E1;0^+_{\rm gs}\rightarrow 1^-)$ of the lowest $1^{-}$ states.
Note that the energy of the $1^{-}_1$ state in $^{142}$Nd was approximately
fitted by adjusting the parameter $v_0$.
The experimental values are taken from Ref. \cite{volz}.}
\begin{center}
\begin{tabular}{c|ddd|ddd} \hline
 &\multicolumn{3}{c|}{$E$ [MeV]}&\multicolumn{3}{c}{$B(E1)\uparrow$ [$e^2$fm$^2$]}\\ \hline 
%Nucleus & RPA & SRPA & Exp. & RPA & SRPA & Exp. \\ \hline
Nucleus & \multicolumn{1}{c}{RPA} & \multicolumn{1}{c}{SRPA} & \multicolumn{1}{c|}{Exp} & \multicolumn{1}{c}{RPA} & \multicolumn{1}{c}{SRPA} & \multicolumn{1}{c}{Exp} \\ \hline
$^{140}$Ce & 7.53 & 3.54 & 3.644 & 0.021 & 0.076 & 0.0217 \\
$^{142}$Nd & 7.36 & 3.45 & 3.424 & 0.010 & 0.074 & 0.0211 \\
$^{144}$Sm & 7.18 & 3.35 & 3.226 & 0.004 & 0.068 & 0.0248 \\ \hline
\end{tabular}
\label{tab1}
\end{center}
\end{table}

In summary, the fragmentation of the dipole strength in
the $N=82$ isotones, $^{142}$Nd, $^{142}$Nd and $^{144}$Sm, was studied using 
the second random-phase approximation (SRPA). 
The SRPA successfully produces the spreading of the giant dipole resonance
and the concentration of the dipole strength  
in the low-energy region, simultaneously.
However, the transition strength of the 
first dipole state was overestimated in the SRPA,
indicating the necessity of a more elaborate treatment for the states
with the two-phonon character.
The calculation based on the extended RPA with
the ground-state correlations is of great interest and currently
under progress.

This work is supported
by Grant-in-Aid for Scientific Research on
Innovative Areas (No. 20105003) and by the Grant-in-Aid for Scientific
Research(B) (No. 21340073).

\end{document}